# Scalable Transfer-Free Fabrication of MoS$_2$/SiO$_2$ Hybrid Nanophotonic Cavity Arrays with Quality Factors Exceeding 4000

Sebastian Hammer [1,2], Hans-Moritz Mangold[1,2], Ariana E. Nguyen[3], Dominic Martinez-Ta[3], Sahar Naghibi Alvillar[3], Ludwig Bartels[3], Hubert J. Krenner[1,2,4,*]

[1] Lehrstuhl für Experimentalphysik 1 and Augsburg Centre for Innovative Technologies (ACIT), Universität Augsburg, Universitätsstraße 1, 86159 Augsburg, Germany

[2] Nanosystems Initiative Munich (NIM), Schellingstraße 4, 80799 München, Germany

[3] Department of Chemistry and Materials Science & Engineering Program, University of California, Riverside, California 92521, USA

[4] Center for Nanoscience (CeNS), Ludwig-Maximilians-Universität München, Geschwister-Scholl-Platz 1, 80539 München, Germany

[*] corresponding author: hubert.krenner@physik.uni-augsburg.de

## ABSTRACT

We report the fully-scalable fabrication of a large array of hybrid molybdenum disulfide (MoS$_2$) - silicon dioxide (SiO$_2$) one-dimensional, free-standing photonic-crystal cavities capable of enhancement of the MoS$_2$ photoluminescence at the narrow cavity resonance. We demonstrate continuous tunability of the cavity resonance wavelength across the entire emission band of MoS$_2$ simply by variation of the photonic crystal periodicity. Device fabrication started by substrate-scale growth of MoS$_2$ using chemical vapor deposition (CVD) on non-birefringent thermal oxide on a silicon wafer; it was followed by lithographic fabrication of a photon crystal nanocavity array on the same substrate at more than 50% yield of functional devices. Our cavities exhibit three dominant modes with measured linewidths less than 0.2 nm, corresponding to quality factors exceeding 4000. All experimental findings are found to be in excellent agreement with finite difference time domain simulations.

## MAIN TEXT

Monolayer transition metal dichalcogenides (TMDs) have recently attracted great interest in the field of photonics because of their distinctive optical and spin properties [1–3]. In contrast to bulk TMD materials, which are indirect bandgap semiconductors, monolayer TMDs are highly optical active due to a direct bandgap ranging between 1 and 2 eV [4,5]. In contrast to top-down fabrication methods like exfoliation [5,6], which yields single TMD flakes of high purity, but does not permit scalability or systematic control of flake thickness and size, chemical vapor deposition (CVD) generates large-scale TMD films of uniform layer thickness and excellent optical and electrical properties[7,8]. Even continuous spatial tuning of the optical bandgap within composite layers of different TMDs has been realized[9,10].

The integration of optically active materials into photonic circuits, for example as light emitters or non-linear elements, requires efficient coupling of TMD excitons to optical fields. This can be dramatically enhanced by nanophotonic resonators. For instance, increased light extraction, signatures of Purcell enhanced emission, and even lasing were observed in systems of TMD monolayers transferred onto two-dimensional GaP-based photonic crystal cavities (PCCs)[11–13] or whispering gallery microdisk resonators[14]. While these studies[11–13] utilized exfoliation of a TMD layer and transfer onto the completed photonic crystal structure, large-scale applications will require direct growth of homogeneous TMD films onto the substrate as an essentially processing step. Thermally grown SiO$_2$ is an excellent candidate substrate for hybrid TMD-PCC devices. It is the most established substrate for CVD growth of TMDs[15] and an essential component of state-of-the-art silicon technology. Moreover, amorphous SiO$_2$ is non-birefringent and exhibits a large optical bandgap of ~9 eV. These properties allow SiO$_2$ to stand out over alternative materials, in particular the commonly-used indirect semiconductor GaP.

Results

Here we demonstrate a sequence of CVD growth of a monolayer MoS$_2$ film followed by fabrication of dozens of free-standing one-dimensional MoS$_2$-SiO$_2$ PCCs on a single substrate without any transfer steps in a fully-scalable approach. Across a broad library of different PCCs fabricated in a single step on the same substrate, we achieve wide linear tunability of the photonic mode spectrum throughout the entire emission band of MoS$_2$, simply by changing the periodicity of the photonic lattice. Both the tuning and the experimentally demonstrated quality ($Q$) factors $Q > 4000$ are confirmed at high fidelity by finite difference time domain (FDTD) simulations.

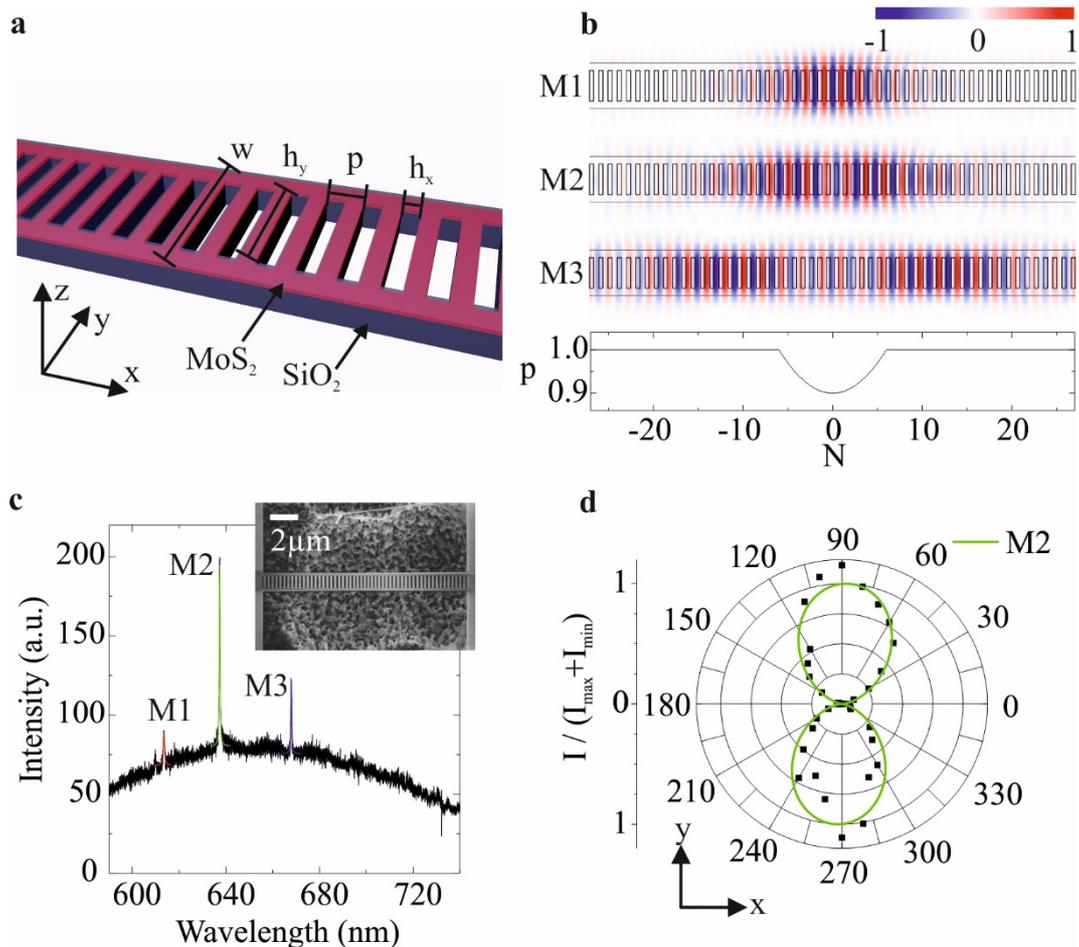

**Figure 1. Structure, design and characterization** – a. Layout of the ladder-type SiO$_2$ (purple) PCC covered on top by a CVD-grown MoS$_2$ monolayer (pink). b. Upper panels: $E_y$ field distribution in the PCC (sketched) of the modes M1, M2, and M3 obtained from FDTD simulation. Lower panel: Period variation in the photonic crystal as a function of the hole position N along the ladder (counted from the center). The defined parabolic perturbation of the period forms the cavity. c. PL spectrum of a typical PCC showing the broad MoS$_2$ emission decorated with three modes, M1, M2 and M3, which are fitted with Lorentzians (colored solid lines). Inset: SEM image of a fabricated PCC. d. Normalized and background corrected peak intensities of the M2 mode (symbols) as a function of the polarization angle and fit (green line) proving TE-polarization.

Due to the relatively low refractive index of SiO$_2$ (~1.46) a ladder-type PCC design is highly desirable[16]. The design used in this study is shown schematically in Figure 1a. This geometry allows for wavelength-scale confinement of the optical modes along the beam direction (*x*) and tight confinement in the perpendicular directions due to total internal reflection. Specifically, our design consists of a 200 nm thick, free-standing, MoS$_2$-SiO$_2$ nanobeam with etched rectangular holes at a base periodicity p. The beam width is *w* = 4.75 *p* and the hole sizes are $h_x$ = 0.5 *p* in *x*-direction and $h_y$ = 0.7 *w* in *y*-direction. As shown in the bottom part of Figure 1b, the period of the first 7 holes in ±*x*-direction, counted from the ladder center, is modulated parabolically from 0.9 *p* to 1.0 *p*. The upper panels of Figure 1b show the $E_y$ field component of the three dominant TE-polarized localized modes, separately calculated by FDTD simulations. As expected[17], the field distribution of the fundamental mode M1 shows a single antinode (maximum) in the center of the PCC. Modes M2 and M3 are the second- and third-order modes of this geometry, respectively.

Our fabrication process starts with CVD growth[15] of MoS$_2$ directly on a 200 nm thermal oxide (SiO$_2$) layer of a Si substrate. After growth, we define the PCC pattern by electron beam lithography. The whole fabrication is described in detail in the methods section. A scanning electron micrograph of a typical free-standing PCC with a period of *p* = 275 nm is shown in the inset in Figure 1c. The total size of a single device, i.e. the area of the undercut region, is approximately 40 µm x 40 µm. In this work, we report on 32 devices fabricated within a rectangular area of 2 mm x 3.5 mm. Closer packing of these relatively large individual devices would in principle allow to integrate more than 5000 devices on a typical 1cm x 1cm substrate. We note, that in our devices the active MoS$_2$ remains only within the rungs of the ladder, precisely the position of the antinodes of the cavity field. This unique feature ensures exclusive enhancement of the light-matter coupling. This is in strong contrast to hybrid devices fabricated by exfoliation and transfer of a 2D material on a prefabricated nanophotonic device[11–13].

Figure 1c shows the room temperature (*T* = 300 K) micro-photoluminescence (µPL) spectrum (see methods section) of a typical PCC of period *p* = 275 nm. The broad background represents the PL emission of unstructured MoS$_2$[4]. In addition, the spectrum shows three sharp lines stemming from the expected localized photonic modes, M1, M2 and M3. Such cavity modes were detected from >50% of all studied PCCs. Lorentzian fits to these peaks (red, green and blue solid lines) reveal *Q*-factors of $Q_{M1}$ = 1000 ($\Delta\lambda$ = 0.62 nm), $Q_{M2}$ = 1700 ($\Delta\lambda = 0.38$ nm) and $Q_{M3}$ = 1570 ($\Delta\lambda = 0.42$ nm), respectively, for this particular device. To confirm the TE-like character of these modes, we perform polarization dependent spectroscopy by placing a half-wave plate and a polarizer into the detection path of the µPL setup. Figure 1d shows the normalized and background corrected intensity of a M2 mode (symbols) as a function of the polarization angle α that is fitted with a sin² function with a period fixed to 180 degrees (green solid line). The polarization direction transverse to the beam (i.e., in *y*-direction) confirms the TE-like mode character. To further quantify the polarized nature of the cavity modes, we evaluate the degree of polarization, *DoP* = ($I_{max}$-$I_{min}$)/( $I_{max}$+$I_{min}$), with $I_{min}$ ($I_{max}$) being the minimum (maximum) intensity. From our data and fit, we obtain a value of *DoP* = 1.000 ± 0.013 that is close to ideal.

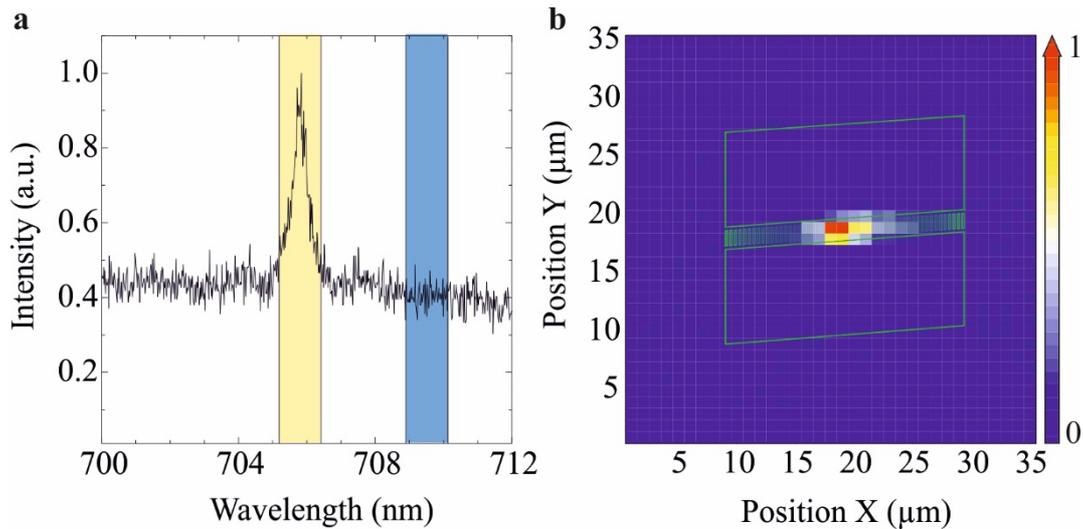

**Figure 2. Spatial confinement** – a. M1 mode of a cavity with period $p$ = 310 nm. The integrated intensity in the spectral region of the cavity mode (left yellow-shaded region) is subtracted from the background (integrated counts of the right blue-shaded region). b. The resulting intensity distribution of the mode M1 shows strong spatial confinement at the center of the PCC (the green outline represents the ladder rungs and the RIE-etched regions defining the outer border of the ladder).

Scanning µPL spectroscopy of a full structure is used to confirm the spatial localization and confinement of a PCC mode. For the data shown in Figure 2a, a total area of 35 µm x 35 µm was scanned at 1µm steps. At each position, a full PL spectrum of the M1 mode was recorded. Figure 2a shows the spectrum recorded at the center of the nanocavity at position ($x/y$) = (19 µm/19 µm). We quantify the enhancement of the M1 emission by integration of the PL emission in the wavelength range of the cavity resonance (marked by the yellow box in Figure 2a) and subtraction of the integrated intensity in a wavelength interval of identical width in close spectral proximity of the mode (marked by the blue box in Figure 2a) for background correction. Performing this procedure at every point of the µPL scan, we are able to build a spatial emission enhancement map, which we normalize to its absolute maximum. Figure 2b shows this emission enhancement map. Superposition of the outline of the etched PCC structure reveals strong emission enhancement at the resonance wavelength of M1 directly in the center of the structure. This finding directly validates the tight spatial localization of the PCC mode.

The PCC resonance wavelengths can be tuned by variation of the crystal period. Figure 3a shows the emission spectra recorded from a series of PCCs with periods tuned from $p$ = 270 nm to 325 nm in steps of $\Delta p$ = 5 nm. The PCC modes exhibit a continuous shift to longer wavelengths as p increases. For short periods $p \leq 285$ nm, modes M1, M2 and M3, marked by red, green and blue arrows, respectively, overlap with the emission band of MoS2. Starting at $p$ = 290 nm (green spectrum), an additional mode M* (black arrow) appears at shorter wavelengths, which also exhibits TE-like character in polarization-dependent measurements. As p increases up to 325 nm, the entire mode spectrum of the PCC continuously shifts across the full $MoS_2$ emission band.

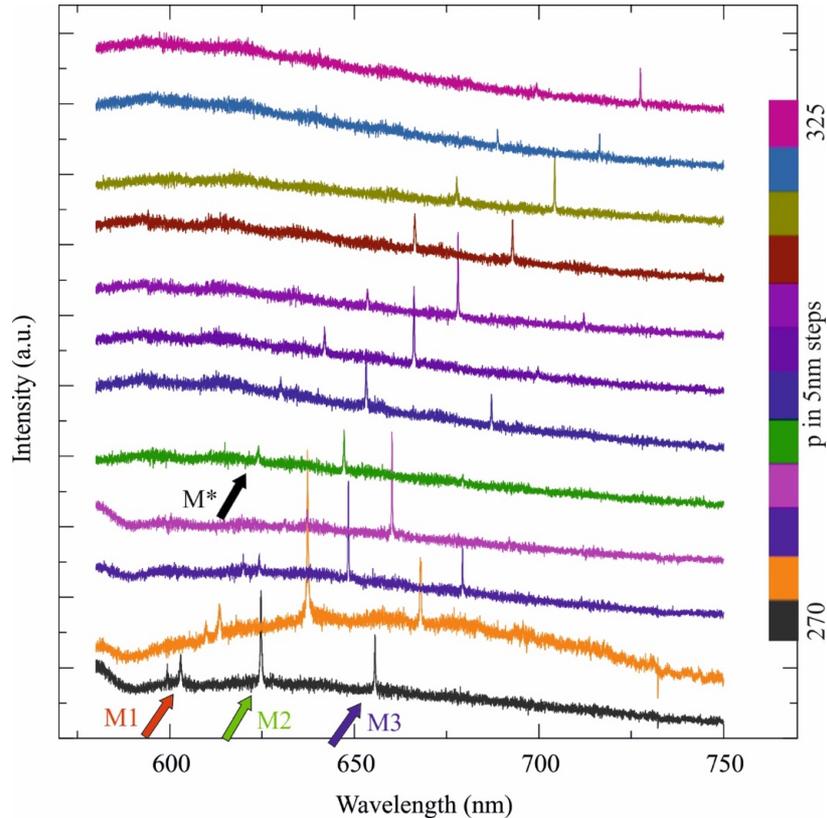

**Figure 3. Geometric tuning** – Spectra of PCCs with periods from 270 nm (bottom) to 325 nm (top). The red (green, blue) arrow marks the M1 (M2, M3) mode. The resonance wavelengths increase linearly with the period. PCCs with p ≥ 290 nm show an additional mode M* (marked by the black arrow).

We analyzed the spectra obtained on each of the PCC structures in detail and extracted both the resonance wavelengths and the *Q*-factors of the PCC modes. For reference, Figure 4a shows a typical PL spectrum of unstructured $MoS_2$. A composite image of a subset of the array of PCC cavities fabricated on a single chip is shown in Figure 4b. Figures 4c and d aggregate the experimentally measured mode wavelengths and Q-factors as a function of the PCC periodicity, respectively. Experimental results are plotted as symbols with the symbol color matched to the mode assignment of Figure 3. The error bars indicate the standard deviation of the mean value obtained by fabrication of more than one PCC at a given periodicity.

Figure 4c reveals clear linear tuning behavior of the resonance wavelengths for all observed cavity modes[16]. Such linear tuning is precisely predicted by our FDTD simulations, which are detailed in the Methods section. The resonance wavelengths of M1 are calculated for $SiO_2$ thicknesses of *d* = 200 nm (the actual $SiO_2$ thickness) and *d* = 300 nm (to account for residual resist left on top of the PCC). The gray area in Figure 4c shows the area where M1 is expected according to the simulations. The measured spectral shift follows faithfully these calculations. Figure 4d shows the *Q*-factor analysis. The measured *Q*-factors are plotted as symbols using the same color-code as in Figures 3 and 4a. Most remarkably, we derive high *Q* > 1000 for almost all PCC modes resolved. However, our experimentally derived values are systematically lower than those observed by Gong and coworkers[16], presumably due to the thinner membranes used in our samples and increased optical absorption loss of $MoS_2$. Our measured *Q*-factors are in excellent agreement with those expected from FDTD modeling (grey shaded area).

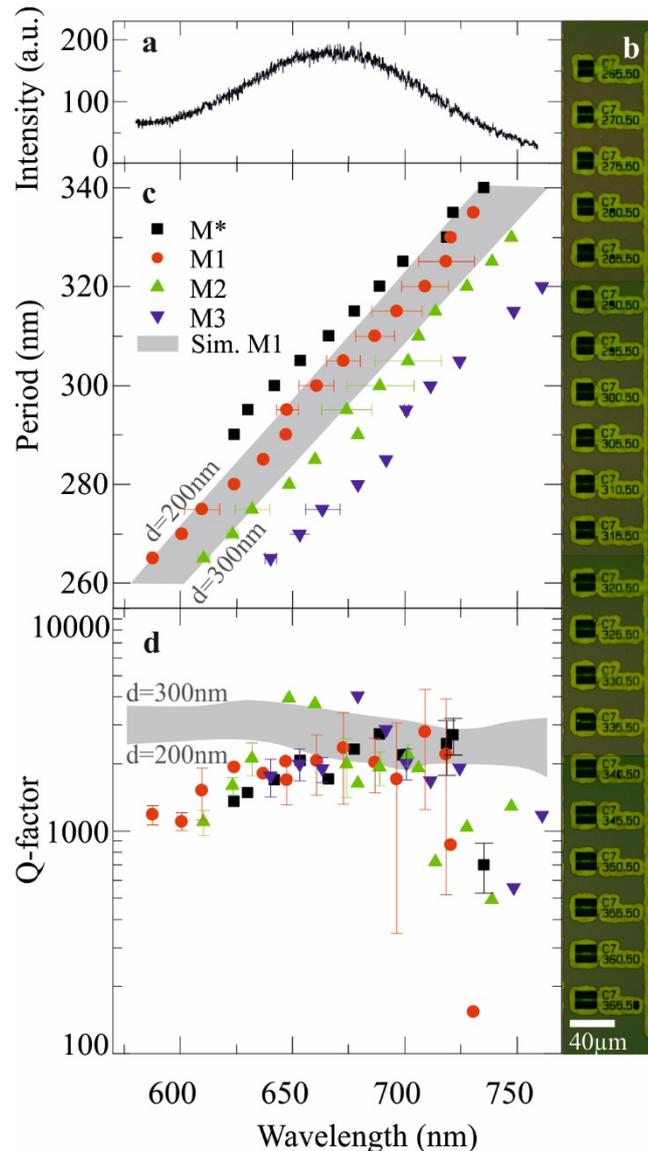

**Figure 4. Tuning analysis and comparison to FDTD simulations** – a. A typical MoS$_2$ PL spectrum. b. Optical microscope image of an array of 21 PCC devices fabricated on a continuous MoS$_2$ film. c. Geometric tuning analysis – The PCC periods within the set of cavities of Figure 3 are plotted as a function of the mode wavelength they support. M1, M2, M3 show excellent agreement with the linear tuning predicted by FDTD simulations for a SiO2 thickness between d = 200 nm and 300 nm (grey shaded area). d. Experimentally observed Q-factor of the three dominant modes as a function of the wavelength the PCC supported. These values agree nicely with FDTD simulation results for M1 (grey shaded area).

## Discussion

We designed and fabricated MoS$_2$-SiO$_2$ PCCs with *Q*-factors > 4000. Simply by tuning the periodicity of the photonic lattice, we are able to tune the PCC mode spectrum across the entire PL emission peak of MoS$_2$. The modes exhibit strong TE polarization and are tightly confined in the cavity region. We show that geometrical tuning of the cavity resonances follows simulations closely and that the measured Q factors are in excellent agreement with our predictions for the ideal structure indicating the high quality throughout dozens of devices fabricated on the same chip – thus, highlighting the efficiency of our fabrication process. Our design can be natively transferred to other TMDs and TMD-based van der Waals heterostructures [18]. In addition, their high *Q*-factor and strong field enhancement are ideally suited to enhance single-photon emission of quantum dot-like emission centers recently observed in the TMD $WSe_2$ [19,20]. The non-birefringent nature of $SiO_2$ is furthermore ideal for future applications in valleytronics. The cavity-enhanced light-matter interactions enable low power gating of the spin and valley degrees of freedom[21,22] using a TMD coupled to an advanced non-polarizing cavity. Finally, the large localized electric fields in the cavity have the potential to offer access to non-linear optical effects such as second harmonic generation. In contrast to a recent report by Fryett and coworkers who used silicon cavities and an exfoliated TMD[23], the large optical bandgap of $SiO_2$ eliminates absorption losses of the visible/near-infrared photoluminescence of many TMDs. Our fully scalable approach is ideally suited for large-scale nanophotonic circuits[24] built from a toolbox of functional design elements[25]. In such circuits the large-area CVD TMD layer will offers multiple functionalities as an optical emitter or a non-linear optical medium[26,27] and a channel layer for associate drive electronics[28]. In particular, simply by adding electrical contacts, the TMD layer can be transformed to an on-chip photodetector[29,30] or a LED [31–33].

METHODS

*CVD growth of $MoS_2$:*

Monolayer $MoS_2$ is grown onto the $SiO_2$/Si substrate via CVD utilizing elemental sulfur and $MoO_3$ powder as precursors[15]. Alumina boats containing the precursors are placed at different positions in a quartz process tube. A molybdenum mesh resting on the edge of the $MoO_3$-containing boat is used to support the substrate. The furnace with the process tube is gradually heated up to 650-700°C. A continuous flow of $N_2$ gas is utilized to transfer sulfur vapor from the boat closer to the furnace edge to the growth substrate near the furnace center. The substrate is held at the indicated temperature for ~10 min for monolayer $MoS_2$ formation and then cooled-down gradually.

Lithographic Device Fabrication: The PCC pattern is written by electron beam lithography into 380 nm thick layer of spin-coated ZEP-520A resist. After development, this pattern is transferred into the $MoS_2$-$SiO_2$ layer by inductively coupled plasma reactive ion etching (ICP-RIE) using a $CHF_3$:$O_2$ chemistry (40sccm:1sccm). Subsequently, and without removing the remaining resist, we release the freestanding PCC nanobeam by etching the underlying Si via an isotropic 2 Torr $XeF_2$ vapor phase process. The resist is not removed prior to optical characterization.

*Optical characterization:*

All samples were characterized at room temperature using conventional µPL spectroscopy. The attenuated (50 µW) beam of a λ = 514 nm diode laser was focused to a ~ 1 µm diffraction limited spot by a 40x, NA = 0.6 microscope objective to locally excite the sample. Emission from the sample is collected by the same objective and spectrally analyzed by a 0.5 m grating monochromator equipped with a liquid $N_2$ cooled Si CCD detector.

*Finite Difference Time Domain Simulations:*

FDTD simulations were performed using the commercially available software package Lumerical FDTD. We used perfectly-matched-layer absorbing boundary conditions of the simulation region. The PCC was excited by electromagnetic pulses of a wide spectral range around the expected cavity resonance that are emitted by three randomly tilted dipole sources that are placed in the center of the cavity. The excitation maximum wavelength was slightly detuned from the expected resonance wavelength to avoid overlapping of excitation and emission signal. Time monitors around the cavity center record the time dependent electromagnetic field from 300 fs to 3000 fs after excitation and calculate the emission spectrum by Fourier transformation.

## ACKNOWLEDGMENTS
This work was funded by Bavaria-California Technology Center (BaCaTeC). S. H., H. M. M. and H. J. K. acknowledge support by Deutsche Forschungsgemeinschaft (DFG) KR3790/2-1 (Emmy Noether Program) and the Cluster of Excellence "Nanosystems Inititative Munich" (NIM). The US National Science Foundation (NSF) supported this work through grant DMR 1609918. Additional support originates from C-SPIN, part of STARnet, a Semiconductor Research Corporation program sponsored by MARCO and DARPA. A.E.N gratefully acknowledge fellowship support under DGE-1326120.

## AUTHOR CONTRIBUTIONS
H. J. K. designed research. S. H. fabricated and characterized photonic crystal cavities. S. H. and H.-M. M. developed fabrication process and FDTD modeling. A. E. N., D. M.-T. and S. N. A. performed CVD growth and initial material characterization. S. H., H. J. K. and L. B. wrote the manuscript. H. J. K. and L. B. supervised study.

## ADDITIONAL INFORMATION
The authors declare no competing financial interest.